\documentclass[english,aps,prl,twocolumn,notitlepage,showpacs,superscriptaddress,nofootinbibfloatfix,nofootinbib]{revtex4-1}
\usepackage[T1]{fontenc}
\usepackage[latin1]{inputenc}
\usepackage{graphicx}
\usepackage{bm}
\usepackage{amssymb}
\usepackage{color}
\usepackage[usenames,dvipsnames]{xcolor}
\usepackage{amsmath}
\usepackage{amstext}
\usepackage{latexsym}
\usepackage[colorlinks=true,citecolor=Sepia,linkcolor=RubineRed,urlcolor=Sepia]{hyperref}
\usepackage{lipsum}
\usepackage{enumitem}

\usepackage{verbatim}

\usepackage{amsfonts}
\usepackage{epsfig}
\usepackage{dsfont}
\usepackage{mathrsfs}
\usepackage{arydshln,leftidx,mathtools}

\begin{document}

\title{Probing Quantum Speed Limits with Ultracold Gases}
\author{Adolfo del Campo}
\affiliation{Department  of  Physics  and  Materials  Science,  University  of  Luxembourg,  L-1511  Luxembourg,  G. D. Luxembourg}
\affiliation{Donostia International Physics Center,  E-20018 San Sebasti\'an, Spain}
\affiliation{IKERBASQUE, Basque Foundation for Science, E-48013 Bilbao, Spain}
\affiliation{Department of Physics, University of Massachusetts, Boston, MA 02125, USA}
\affiliation{Theory Division, Los Alamos National Laboratory, MS-B213, Los Alamos, NM 87545, USA}

\def\q{{\bf q}}

\def\G{\Gamma}
\def\L{\Lambda}
\def\la{\lambda}
\def\g{\gamma}
\def\al{\alpha}
\def\s{\sigma}
\def\e{\epsilon}
\def\k{\kappa}
\def\ve{\varepsilon}
\def\l{\left}
\def\r{\right}
\def\te{\mbox{e}}
\def\d{{\rm d}}
\def\t{{\rm t}}
\def\K{{\rm K}}
\def\N{{\rm N}}
\def\H{{\rm H}}
\def\la{\langle}
\def\ra{\rangle}
\def\om{\omega}
\def\Om{\Omega}
\def\vep{\varepsilon}
\def\wh{\widehat}
\def\tr{{\rm Tr}}
\def\da{\dagger}
\def\iz{\left}
\def\zi{\right}
\newcommand{\beq}{\begin{equation}}
\newcommand{\eeq}{\end{equation}}
\newcommand{\beqa}{\begin{eqnarray}}
\newcommand{\eeqa}{\end{eqnarray}}
\newcommand{\intf}{\int_{-\infty}^\infty}
\newcommand{\into}{\int_0^\infty}

\begin{abstract}
 Quantum Speed Limits (QSLs) rule the minimum time  for a quantum state to evolve into a distinguishable state in an arbitrary physical process. These fundamental results constrain  a notion of distance travelled by the quantum state, known as the Bures angle, in terms of the speed of evolution set  by nonadiabatic energy fluctuations. We theoretically propose how to measure  QSLs in an ultracold quantum gas confined in a time-dependent harmonic trap. In this highly-dimensional system of continuous variables, quantum tomography is prohibited. Yet, QSLs can be probed whenever the dynamics is  self-similar by measuring as a function of time the cloud size of the ultracold gas. This  makes possible to determine the Bures angle and energy fluctuations, as we discuss  for  various  ultracold atomic systems. 

\end{abstract}

\maketitle

The time-energy uncertainty relation is a fundamental result in quantum physics relating characteristic times to the inverse of energy fluctuations \cite{Pfeifer95,Busch08}. This seminal result goes back to Mandelstam and Tamm who established it rigorously in 1945 \cite{MT45}. Its modern formulation relies on quantum speed limits (QSLs) that bound the minimum time for a physical process to unfold in terms of energy fluctuations. QSLs render quantum dynamics with a geometric interpretation in which the quantum state of a system evolves in time by sweeping a distance in Hilbert space \cite{DeffnerCampbell17}.  Thus, QSLs involve the notions of speed and distance in Hilbert space.
Quantifying the distance between the initial and time-evolving quantum states requires estimating state overlaps, which is challenging, if not unfeasible, for many-particle systems with continuous variables.
Different norms of the generator of evolution provide upper bounds to the speed at which this distance is traversed. 
Apart from the  standard deviation of the energy  \cite{MT45,Messiah61,Bhattacharyya83,AA90,Uhlmann92,Pfeifer93,Busch08}, the mean energy above the ground state has been widely used after the QSL introduced by Margolus and Levitin \cite{ML98,Giovannetti11}. In addition, other moments of the Hamiltonian can  be used to upper bound  the speed of evolution \cite{Zych06,Margolus11}, and, in certain settings, other notions of speed based on work fluctuations have been shown to be dominant \cite{Funo17}.

By now, QSLs are established in open quantum systems \cite{Taddei13,delcampo13,DeffnerLutz13,Campaioli2019} and stochastic evolutions under continuous quantum measurements \cite{Luispe19,garcapintos19b}. Indeed, it is at present understood that speed limits are not restricted to the quantum domain, and can be formulated universally using the tools of information geometry \cite{Amari16}. The derivation of speed limits in classical dynamics and stochastic thermodynamics constitute a compelling advance to this end \cite{Shanahan18,Okuyama18,Shiraishi18}. 
The notion of distinguishability in classical and quantum systems is however fundamentally different. 
In the quantum domain, the default notion relies on the Bures angle \cite{Wootters81,Uhlmann92}. Alternatively, other measures such as the Wigner-Yanase information \cite{Pires16} and the generalized Bloch angle \cite{Campaioli18} have been explored.

In spite of the fundamental nature of QSLs, there is currently a lack of experimental studies probing them.  In this work we propose the experimental study of QSLs with many-body systems of trapped ultracold atoms by measuring the mean atomic cloud size as a function of the evolution time. We show that for scale-invariant many-body systems, the  Mandelstam-Tamm QSL can be probed,  given that  the Bures angle as well as the nonadiabatic energy fluctuations can be determined from the mean atomic cloud size, which is an experimentally measurable quantity.

{\it Geometry of quantum dynamics and  QSL.---}
The degree to which two pure quantum states resemble each other is captured by the absolute square value of their overlap, i.e., their fidelity. Consider an  initial quantum state $|\Psi(0)\ra$ and its time-evolution after a time $t$ denoted by $|\Psi(t)\ra=U(t,0)|\Psi(0)\ra$, where $U(t,0)$ is the unitary time-evolution operator generated  by the system Hamiltonian  dynamics assuming that the system is isolated from the external environment.
The fidelity $F(t)=|\la\Psi(0)|U|\Psi(0)\ra|^2$  gives  the survival probability of the initial state after a time $t$ of evolution. 
 A notion of distance between quantum states is provided by the Bures angle \cite{Wootters81,Uhlmann92}. In particular, the Bures angle  between the initial and the time-dependent states  
 reads
 \beqa
 \label{Bures}
 \mathcal{L}(t)= \mathcal{L}(|\Psi(0)\ra,U|\Psi(0)\ra)={\rm arccos}\sqrt{F(t)}.
 \eeqa
The Bures angle swept during the evolution is upper bounded in terms of the quantum Fisher information $I_Q$,
 \beqa
  \mathcal{L}(\tau)\leq \int_0^\tau ds\sqrt{I_Q(t)/4}.
  \label{MTbound}
 \eeqa 
Under unitary evolution, the quantum Fisher information is proportional to the energy variance, i.e., 
\beqa
I_Q(t)=
\frac{4}{\hbar^2}[\la\Psi(t)| H(t)^2|\Psi(t)\ra- \la\Psi(t)| H(t)|\Psi(t)\ra^2].\nonumber
\eeqa
This results in the Mandelstam-Tamm QSL \cite{MT45,Uhlmann92,Pfeifer93,Busch08}
\beqa
\tau\geq\tau_{\rm QSL}=\frac{ \hbar \mathcal{L}(\tau)}{\overline{\Delta H}},
\eeqa
where the mean energy dispersion reads
\beqa
\overline{\Delta H}=\frac{1}{\tau}\int_0^\tau dt \sqrt{{\rm var}_{\rho(t)}[H(t)]}.
\eeqa
This QSL can be used to characterize a given evolution.
To this end, we introduce the difference between the integrated nonadiabatic standard deviation of the energy and the Bures angle
\beqa
\delta \mathcal{L}(\tau)=\frac{1}{\hbar}\int_0^\tau dt \sqrt{{\rm var}_{\rho(t)}[H(t)]}- \mathcal{L}(\tau)\geq 0.
\eeqa
The  first term in the rhs,   $\gamma(\tau)=\tau\overline{\Delta H}$, represents the length  of  the path followed during the evolution in  projective Hilbert space from $\Psi(0)$ to $\Psi(\tau)$ \cite{AA90,Andersson14}, 
\beqa
\gamma(\tau)=\int_0^\tau dt\sqrt{\la d_t\Psi(t)|[1-P(t)]|d_t\Psi(t)\ra},
\eeqa
 with $P(t)=|\Psi(t)\ra\la \Psi(t)|$. This length cannot be smaller than the actual geodesic $ \mathcal{L}(\tau)$ between the two states, i.e., the distance defined by Eq. (\ref{Bures}).
Thus, the quantity  $\delta \mathcal{L}(\tau)$ quantifies the extent to which a given evolution saturates the QSL. Said differently, when $\delta \mathcal{L}(\tau)$ vanishes, 
the evolution takes place at the maximum speed allowed by the Mandelstam-Tamm bound at all times during the considered time interval $[0,\tau]$.

{\it Trapped ultracold gases with self-similar dynamics.---}
We next show how  to determine the QSL in ultracold atomic gases. Consider the family of  time-dependent Hamiltonians
\begin{eqnarray}
 H(t)=\sum_{i=1}^{\N}\left[\frac{\vec{p}_i\,^2}{2m}+\frac{1}{2}m\omega(t)^2 \vec{r}_i\,^{2}\right]+\sum_{i<j}V(\vec{r}_i-\vec{r}_j),
 \label{hscale}
\end{eqnarray}
describing $\N$ particles in a harmonic trap. Particles interact with each other through a homogeneous pairwise potential  fulfilling $V(\lambda \vec{r})=\lambda^{-2}V(\vec{r})$.  Thanks to this scaling property, the dynamics is self-similar, i.e., scale invariant \cite{Castin04,Gritsev10,delcampo11}, a familiar feature in  Bose-Einstein condensates \cite{Kagan96,CastinDum96}.
An energy eigenstate $\Psi(0)$ of the Hamiltonian at $t=0$  with eigenvalue $E(0)$ evolves into
\beqa
\label{psit}
\Psi\left(t\right)&=&
\frac{1}{b^{\frac{D\N}{2}}}\exp\left[i\frac{m\dot{b}}{2\hbar b}\sum_{i=1}^\N \vec{r}_i\,^2-i\int_{0}^t\frac{E(0)}{\hbar b(t')^2}dt'\right]\nonumber\\
& &\times \Psi\left(\frac{\vec{r}_1}{b},\dots,\frac{\vec{r}_\N}{b},t=0\!\right)\,,
\eeqa
where  $D$ denotes the spatial dimension and $b(t)$ is the scaling factor that determines the atomic cloud size. The specific  time-dependence of the latter following an arbitrary modulation of the trapping frequency 
$\om(t)$ can be found by solving the Ermakov equation, 
$ \ddot{b}+\om(t)^2b=\om_0^2/b^{3}$, 
with the  boundary conditions $b(0)=1$ and $\dot{b}(0)=0$, as $\Psi(0)$ is assumed to be stationary for $t<0$.

While the scale invariant dynamics facilitates the description of the time evolution, the study of QSL remains hindered by the requirement to compute the Bures angle.
Direct measurement of the overlap between quantum states is generally difficult in many-body systems, in particular, in the case of continuous variables. 
However, we shall show  that for a low-energy state in a variety of systems, the Bures angle can be expressed solely in terms of the scaling factor, which is an experimentally measurable quantity.

To relate the  Bures angle to the ultracold-gas cloud size, we first
consider the system Hamiltonian in the absence of a trap
\beqa
 H_{\rm free}=\sum_{i=1}^{\N}\frac{\vec{p}_i\,^2}{2m}+\sum_{i<j}V(\vec{r}_i-\vec{r}_j)\ ,
 \label{hscale}
 \eeqa
 and let  $\psi_\nu$ be an energy eigenstate satisfying $H_{\rm free}\psi_\nu=\varepsilon_\nu \psi_\nu$,  that is also a homogeneous function 
\beqa
\psi_\nu\left(\lambda \vec{r}_1,\dots,\lambda \vec{r}_\N\right)=\lambda^\nu\psi_\nu(\vec{r}_1,\dots,\vec{r}_\N),
\eeqa
i.e., it is an eigenstate of the dilatation operator $\sum_{i=1}^{\N} \vec{r}_i\cdot \nabla_{\vec{r}_i}\psi_\nu=\nu\psi_\nu$.
Then,  the ground-state wavefunction of the  Hamiltonian $H(0)= H_{\rm free}+\frac{1}{2}m\omega_0^2 \sum_{i=1}^\N\vec{r}_i\,^{2}$ in equation (\ref{hscale}) reads
\beqa
\Psi_0(\vec{r}_1,\dots,\vec{r}_\N)=c_0e^{-\frac{m\om_0}{2\hbar}\sum_{i=1}^\N \vec{r}_i\,^2}\psi_\nu(\vec{r}_1,\dots,\vec{r}_\N),
\eeqa
where $c_0$ is a normalization constant and the energy eigenvalue  is 
$
E(0)=\varepsilon_\nu+\hbar\om_0\left(\nu+\frac{D\N }{2}\right).
$
This relation between eigenstates in the presence and absence of a trap is realized in a variety of systems \cite{SM,Olshanii98,WernerCastin06,Beau16}.
It holds (trivially) for  the ground-state of the single-particle harmonic oscillator. It also applies to the ground state of one-dimensional many-body systems such as the free Bose gas, 
a polarized free Fermi gas, the Tonks-Giraradeau gas and the Calogero-Sutherland gas \cite{delcampo20}.  In three spatial dimensions it describes a family of states of 
the unitary Fermi gas \cite{tan2004short,Castin12}.

Upon varying $\om(t)$, the self-similar evolution (\ref{psit})  yields
\beqa
\Psi_0(t)=\frac{c_0e^{i\alpha_t}}{b^{\nu+\frac{D\N}{2}}}e^{-\frac{m\om_0}{2\hbar}\left(1-i\frac{\dot{b}}{\om_0 b}\right)\sum_{i=1}^\N \vec{r}_i\,^2}\psi_\nu(0),
\eeqa
with $\alpha_t=-\int_{0}^t\frac{E_m(0)}{\hbar b(t')^2}dt'$ being a dynamical phase.
We find that the overlap between $\Psi_0(0)$ and its time evolution at time $t$ equals
 \beqa
 \la  \Psi_0|U(t,0)| \Psi_0\ra=e^{i\alpha_t}\left[\frac{b}{2}\left(1+\frac{1}{b^2}-i\frac{\dot{b}}{\om_0 b}\right)\right]^{-\sigma^2}\!\!\!\!\!\!\!,
 \eeqa
where  $\sigma^2=\nu+\frac{D\N}{2}$  in spatial dimension $D$.

Its absolute value is the square root of the fidelity used to define the Bures angle
  \beqa
  \label{Fidelityt}
\sqrt{F(t)}=\left[\frac{b^2}{4}\left[\left(1+\frac{1}{b^2}\right)^2+\left(\frac{\dot{b}}{\om_0 b}\right)^2\right]\right]^{-\frac{\sigma^2}{2}}.
 \eeqa
We further note that
\beqa
\sigma^2=\frac{E(0)}{\hbar\om_0}=\frac{1}{x_0^2}\la  \Psi_0|\sum_{i=1}^\N \vec{r}_i\,^2| \Psi_0\ra,
\eeqa
 which means that $\sigma$ is the initial size of the cloud formed by the ultracold gas in units of $x_0=\sqrt{\hbar/(m\om_0)}$, that can be experimentally measured. 
 As a result, the Bures angle swept during a time of evolution $t$ can be determined from the time-dependent scaling factor $b(t)$.
Remarkably, the expression for the fidelity (\ref{Fidelityt}) holds for a variety of harmonically trapped quantum systems when $|\Psi_0\ra$ is chosen to be the ground state with energy $E(0)$. 
See  \cite{SM} for the derivation of the values of $\sigma^2$ summarized here:
For a $D$-dimensional quantum oscillator  ($\N=1$), $\sigma^2=\frac{D}{2}$. For a trapped noninteracting  Bose gas ($V=0$),  $\sigma^2=\frac{\N D}{2}$, while for a spin-polarized Fermi gas, $\sigma^2=\frac{\N^2 D}{2}$.
For bosonic systems in one spatial dimension $D=1$, whenever $V$ describes hard-core interactions one recovers  the Tonks-Girardeau gas \cite{Girardeau01,MinguzziGangardt05}, experimentally realized in \cite{Kinoshita04,Paredes04,Palzer09,Jacqmin11}. In this case, $\sigma^2=\frac{\N^2}{2}$, which matches the result of a one-dimensional spin-polarized Fermi gas as a result of the Bose-Fermi mapping \cite{Girardeau60,Zurn12,Wenz13}. 
For the rational Calogero-Sutherland model in which $V$ represents inverse-square pairwise interactions of strength $\lambda$  \cite{Calogero71,Sutherland71,Sutherland98}, $\sigma^2=\N[1+\lambda(\N-1)]/2$.
In addition, for a unitary Fermi gas in three spatial dimensions \cite{Castin04,Castin12,Deng16,Deng18,Deng18sci}, one can make use of  the general expression $\sigma^2=E(0)/(\hbar\om_0)$ \cite{Beau20}.

The vanishing of the fidelity (\ref{Fidelityt}) for $t>0$ in many-body systems can be considered a manifestation of  the orthogonality catastrophe \cite{Anderson67}, encoded in the dependence of $\sigma^2$ on the particle number $\N$. In particular, the scaling $\sigma^2\propto \N^2$ is not only shared by spin-polarized fermions and hard-core bosons \cite{delcampo11long,Fogarty20}, but as well by the Calogero-Sutherland gas \cite{delcampo16,Ares18}.

 Apart from the Bures angle, the study of QSL requires knowledge of the speed of evolution.
Under scale-invariant dynamics generated by the time-dependent Hamiltonian (\ref{hscale}), the  energy variance  in a state (\ref{psit}) is 
\beqa
{\rm var}_{\rho(t)}[H(t)]=\hbar^2\om(t)^2\sigma^2
 \left[\left(Q^*\right)^2-1\right].
\eeqa
Here, the nonadiabatic factor $Q^{\ast}(t)$  is given by
\beqa\label{Qstar}
Q^{\ast}(t)=\frac{\om_0}{\om(t)} \left[\frac{1}{2b(t)^{2}}+\frac{\om(t)^2 b(t)^2}{2\om_0^2}+\frac{\dot{b}(t)^2}{2\om_0^2}\right],
\eeqa
and accounts for the amount of energy excitations over the adiabatic dynamics. Indeed, 
$Q^{\ast}(t)=\la H(t)\ra/\la H(t)\ra_{\rm ad}$ is the ratio between 
the nonadiabatic mean energy $\la H(t)\ra$ and the mean energy under adiabatic driving $\la H(t)\ra_{\rm ad}=\la H(0)\ra \om(t)/\om_0$ \cite{Jaramillo16,Abah17,Beau20,AbahLutz18}. 
Thus, the integrated  mean energy dispersion is  given by
\beqa
\gamma(\tau)=\sigma\int_0^\tau dt \sqrt{\om(t)^2 \left[\left(Q(t)^*\right)^2-1\right]}
\eeqa
and, together with Eq. (\ref{Fidelityt}) it determines the QSL in Eq. (\ref{MTbound}). 
QSLs can  thus be probed by determining $b(t)$ from time-of-flight imaging data, using a similar analysis to that used for determination of the nonadiabatic scaling factor $Q^*$ reported in \cite{Deng18sci,SM}.

{\it  Generic expansion versus a shortcut.---}
%
\begin{figure}[t]
\includegraphics[width=1\linewidth]{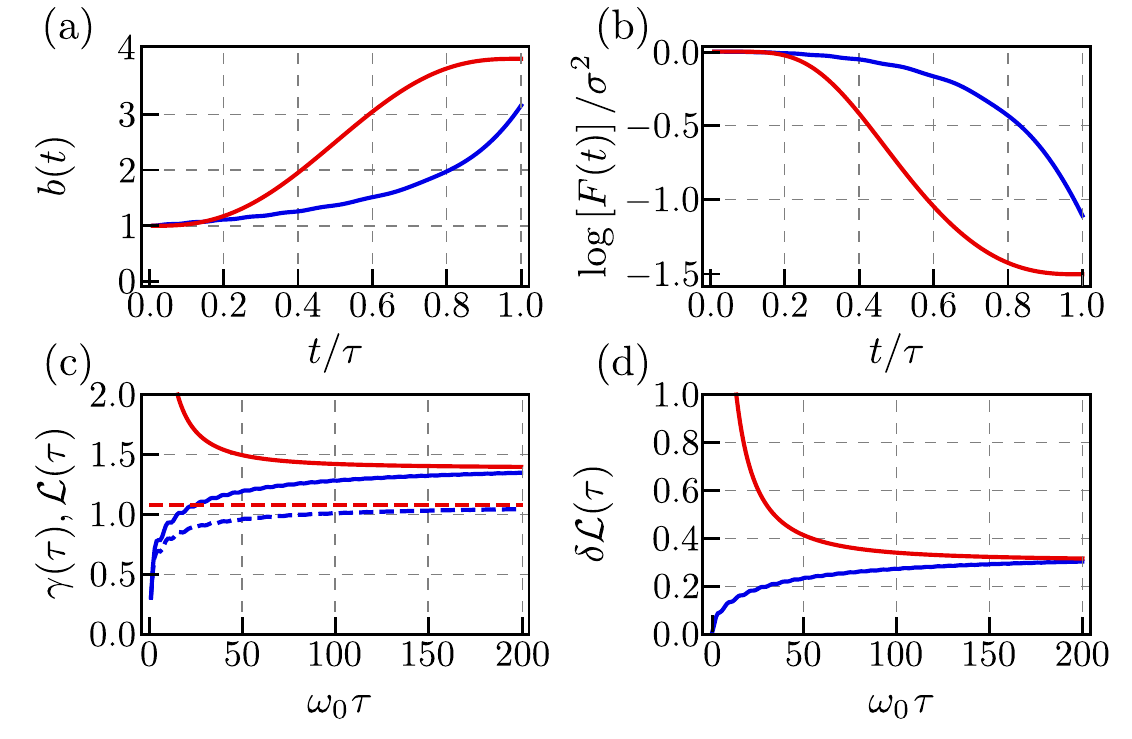}
\vspace{-0.5cm}
\caption{\label{Fig1QSL} {\bf  QSL for an expansion induced by a linear frequency ramp and a shortcut to adiabaticity. } (a) Scaling factor and (b)  logarithmic fidelity as a function of time for a four-fold expansion with $\tau=10/\om_0$ for a linear ramp (blue) and a STA (red). Orthogonality catastrophe is encoded in the constant $\sigma^2$ which captures the dependence on the system size. (c) Path length $\gamma(\tau)$ (solid) in Hilbert space lower bounded by the geodesic $\mathcal{L}(\tau)$ (dashed) with $\N=1$. (d)  While the  excess Bures angle $\delta\mathcal{L}$  increases for a linear ramp as  the adiabatic limit is approached, the converse is true for the STA.}
\end{figure}
We next analyze the nonadiabatic expansion resulting from varying the trap frequency from an initial value $\om_0$ to a final one $\om_F<\om_0$ in  an expansion time $\tau$. 
We first consider a linear ramp $\om(t)=\om_0+(\om_F-\om_0)t/\tau$, for which the scaling factor $b(t)$ is determined by solving numerically the Ermakov equation. We compare it with a shortcut to adiabaticity (STA) designed  by reverse engineering the scale-invariant dynamics \cite{Chen10,delcampo11,SM}. The latter is based on fixing first a trajectory of the scaling factor  $b(t)$ interpolating between the boundary conditions $b(0)=1$ and $b(\tau)=\sqrt{\om_0/\om_F}$, the later being  the target adiabatic value obtained by setting $\ddot{b}\approx 0$ in the Ermakov equation. For the initial and final states to be nonstationary, Eq. (\ref{psit}) imposes that $\dot{b}(0)=\dot{b}(\tau)=0$. The polynomial ansatz $b(t)=1+10(t/\tau)^3(b(\tau)-1)-15(t/\tau)^4(b(\tau)-1)+6(t/\tau)^5(b(\tau)-1)$ is  thus chosen, satisfying as well $\ddot{b}(0)=\ddot{b}(\tau)=0$, and the trap frequency $\om(t)$ is determined from the Ermakov equation as $\om(t)^2=\om_0/b^4-\ddot{b}/b$.

Fig \ref{Fig1QSL} shows the scaling factor as a function of time in a fast expansion. In the linear ramp, $b(t)$ does not reach the adiabatic value corresponding to the final frequency, $b(\tau)= \sqrt{\om_0/\om_F}=4$, while  for the STA it interpolates for any $\tau$ between the initial and target configuration. Given the scaling factor $b(t)$, we obtain the fidelity along the process using Eq. (\ref{Fidelityt}), and show its  monotonic decay in both cases. Knowledge of the scaling factor also allows us to determine the integrated energy dispersion that sets the length $\gamma(\tau)$ of the path travel in Hilbert state, and which is lower bounded by the geodesic $\mathcal{L}$. This demonstrates that the QSL is fulfilled during the dynamics, in any process. Moreover, the difference  between $\gamma(\tau)$ and $\mathcal{L}(\tau)$ shows the extent to which the evolution saturates  the QSL.  For an arbitrary expansion time $\tau$, a  linear ramp follows more closely the QSL than the STA, but yields lower values of  $\gamma(\tau)$ and $\mathcal{L}(\tau)$. Indeed,  for fast expansions both quantities vanish with a linear ramp, while a STA involves large deviations from QSL  and has a $\mathcal{L}(\tau)$ independent of $\tau$. For slow expansions with $\om_0\tau\gg1$, both protocols behave alike. In the adiabatic limit, $\delta \mathcal{L}$ is still finite as we next show.

{\it Example 2. Adiabatic and Transitionless Quantum driving.---}
Counterdiabatic or  transitionless quantum driving (TQD) is a technique that enforces the evolution of the state along a prescribed adiabatic trajectory \cite{Demirplak03,Demirplak05,Berry09}. To this end, an auxiliary control field is introduced to assist the dynamics and enforce parallel transport.  Takahashi has shown that TQD solves the quantum brachistochrone  \cite{Takahashi13}, this is, the variational problem of minimizing the evolution time between and initial and a final state under fixed energy variance \cite{Carlini06}. We analyze to what extent the resulting evolution minimizes $\delta \mathcal{L}(\tau)$.

For the time-dependent Hamiltonian (\ref{hscale}), the adiabatic evolution can be obtained from (\ref{psit}) by considering the adiabatic scaling factor $b(t)=\sqrt{\om_0/\om(t)}$. Using this expression in  (\ref{psit}) while setting $\dot{b}\approx 0$, yields
\beqa
\label{psitad}
\Psi\left(t\right)&=&
\frac{e^{i\alpha_t}}{b^{\frac{D\N}{2}}}\
\Psi\left(\frac{\vec{r}_1}{b},\dots,\frac{\vec{r}_\N}{b},t=0\!\right),
\eeqa
The auxiliary control field that assists the dynamics along this adiabatic  trajectory is given by \cite{Jarzynski13,delcampo13cd}
 \beqa
H_1(t)=\frac{\dot{b}}{b}C=\frac{\dot{b}}{b}\frac{1}{2}\sum_{i=1}^{\N}\left\{\vec{r}_i,\vec{p}_i\right\},
\eeqa 
where $C$ is the squeezing operator.
Thus, the evolution (\ref{psitad}) is the exact solution of the many-body time-dependent Sch\"odinger equation with the Hamiltonian $H_T=H(t)+H_1(t)$.
In this case, the energy variance reduces to the second-moment of the auxiliary term 
$\Delta  H_T^2=\la  H_1^2\ra$ \cite{Campo2012a,Bukov19}.
The nonadiabatic energy variance can then be written as  \cite{SM}
\beqa
\Delta  H_T^2=\left(\frac{\dot{b}}{b}\right)^2\la C^2(t)\ra=\left(\frac{\dot{b}}{b}\right)^2\hbar^2\sigma^2.
\eeqa
The Mandelstam-Tamm upper bound to the speed of evolution is thus governed by the second moment of the squeezing operator, which is time-independent.
Explicit integration yields the path length travelled $\gamma(\tau)$
\beqa
\label{gammacd}
\int_0^\tau dt \frac{\Delta  H_T(t)}{\hbar}=\sigma\alpha\log b(\tau)=\log\left(\frac{\om(\tau)}{\om_0}\right)^{-\alpha\frac{\sigma}{2}}\!\!,
\eeqa
assuming $b(t)$, and thus $\om(t)$, to be monotonic. In this case, $\alpha={\rm sgn}(\dot{b})$ reduces to $+1$ in an expansion and to $-1$ in a compression.
%
\begin{figure}[t]
\begin{center}
\includegraphics[width=0.8\linewidth]{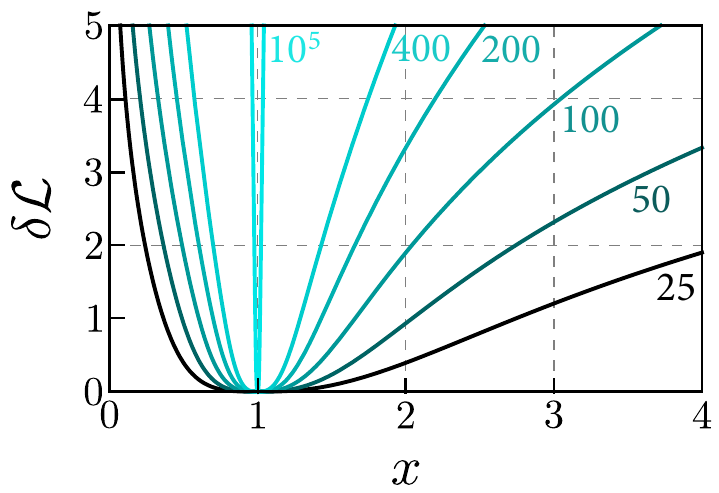}
\vspace{-0.5cm}
\end{center}
\caption{\label{Fig2QSL} {\bf  Excess Bures angle  under adiabatic evolution and TQD.} $\delta \mathcal{L}(\tau)$ is shown for different values of $\sigma^2$, increasing from bottom to top. The the Mandelstam-Tamm QSL is only saturated when the ratio $x=\om(\tau)/\om_0$ approaches unity, as $\delta \mathcal{L}(\tau)$  vanishes. Many particle effects  increase $\delta \mathcal{L}(\tau)$ hindering the driving at the QSL.}
\end{figure}
Under TQD,  the Bures angle is set by the overlap between the initial eigenstate of $H(0)$ and its adiabatic continuation (\ref{psitad}) at time $t$, 
\beqa
F(\tau)=
\left[\frac{\om_0}{4\om(\tau)}\left(1+\frac{\om(\tau)}{\om_0}\right)^2\right]^{-\sigma^2}.
\eeqa
In this case, the  excess Bures angle reads
\beqa
\label{dLad}
\delta \mathcal{L}(\tau)=-\alpha\frac{\sigma}{2}\log x-\arccos\left[\left(\frac{1+x}{2\sqrt{x}}\right)^{-\sigma^2}\right],
\eeqa
where $x=\om(\tau)/\om_0$ and is shown in Fig. \ref{Fig2QSL}.
For small expansions and compressions, 
\beqa
\label{dLexp}
\delta \mathcal{L}(\tau)=\alpha\sigma(1-x)+\frac{\alpha\sigma}{2}(1-x)^2+\mathcal{O}[(x-1)^3].
\eeqa
As a result, the excess Bures angle $\delta \mathcal{L}(\tau)$ remains finite for any  ratio between the final and the initial frequency $x=\om(\tau)/\om_0\neq 1$. The characteristic range of the frequency ratio in which $\delta \mathcal{L}(\tau)$ is negligible is set by the inverse of ${\sigma}$.   It is thus reduced for many particle systems as the  particle number $\N$ and the spatial dimension $D$ are increased.  The results (\ref{gammacd})-(\ref{dLexp}) not only describe TQD but also apply in the adiabatic limit \cite{SM}. Indeed, Eq. (\ref{dLad}) for $x=1/16$ yields $\delta \mathcal{L}(\tau)\approx0.305$, the asymtoptic value in Fig. \ref{Fig1QSL} for large $\tau$. In short, the QSL defined in terms of the Bures angle is never saturated, as the geodesic $ \mathcal{L}(\tau)$ cannot be accessed with the considered dynamics. The reachable geodesic for $\gamma(\tau)$ is  the one defined on the manifold of adiabatically accessible states, as conjectured in \cite{Bukov19} for arbitrary dynamics.

{\it Summary and conclusions.---}
We have demonstrated that QSLs can be  probed in ultracold atom experiments characterized by self-similar dynamics.
Our proposal relies on measuring  the size of the atomic cloud in a given process, such an expansion or compression driven by a modulation of the trap frequency. 
The scaling factor can be determined by imaging the cloud size via standard time-of flight measurements or non-destructive Faraday imaging \cite{Miroslav13}, among other approaches.
From it,  one can determine the distance travelled by the quantum state of the system in Hilbert space (Bures angle) during the evolution. This approach circumvents the need for quantum state tomography of the many-body state of a continuous variable system.
In addition, the scaling factor also determines the Mandelstam-Tamm quantum speed of evolution, that equals the time-average of the energy dispersion. 
Their knowledge allows one to quantify the extent to which a given evolution saturates the QSL, paving the way to identifying  time-optimal protocols \cite{Caneva09}, as we have discussed in the context of fast control by shortcuts to adiabaticity.
Our proposal is amenable to experimental studies with trapped ultracold atomic clouds in three spatial dimensions (e.g. \cite{Schaff10}). Similarly, it can be applied to ultracold gases in tight waveguides with an axial harmonic confinement \cite{Wenz13}. In an isotropic setting, it can further be implemented at strong coupling using a unitary Fermi gas \cite{Deng16}. 
These results pave the way to the experimental study of the time-energy uncertainty relation and QSLs  in many-body quantum systems, and their relation to the orthogonality catastrophe.

{\it Acknowledgements.---} It is a pleasure to acknowledge discussions with Andrea Alberti, Tommaso Calarco,  L\'eonce Dupays, \'I\~nigo L. Egusquiza,  Fernando J. G\'omez-Ruiz, Dries Sels, Kazutaka Takahashi 
and Zhenyu Xu. 
This work was supported by PID2019-109007GA-I00.

\bibliography{QSLultracold_lib}

\newpage

\clearpage

\widetext
\begin{center}
\textbf{{\large Supplemental Material---Probing Quantum Speed Limits with Ultracold Gases}}
\end{center}

\setcounter{equation}{0} \setcounter{figure}{0} \setcounter{table}{0}
\makeatletter
\renewcommand{\theequation}{S\arabic{equation}} \renewcommand{\thefigure}{S%
\arabic{figure}} \renewcommand{\bibnumfmt}[1]{[#1]} \renewcommand{%
\citenumfont}[1]{#1}

\tableofcontents

\section{Quantum Many-body States of Trapped Systems: Values of $\nu$ and $\sigma^2$}

{\it Tonks-Girardeau gas.---}
Ultracold bosons confined in a tight-waveguide are well-described by the one-dimensional Lieb-Liniger gas with contact interactions $V(x_i-x_j)=g\delta(x_i-x_j)$ with coupling strength $g$ \cite{Olshanii98}.
The limit in which $g\rightarrow\infty$ describes hard-core bosons and is known as the Tonks-Girardeau gas \cite{Girardeau60}.
As hard-core interactions mimic  Pauli exclusion, it is possible to relate the wavefunction of a Tonks-Girardeau gas $\Psi_{TG}$ to that of an ideal (spin-polarized) Fermi gas in one spatial dimension $\Psi_{F}$. Specifically, the Bose-Fermi mapping reads \cite{Girardeau60}
\beqa
\Psi_{TG}(x_1,\dots,x_\N)=\prod_{i<j}{\rm sgn}(x_{i}-x_{j})\Psi_{F}(x_1,\dots,x_\N),
\label{BFM}
\eeqa
where ${\rm sgn}(x)$ is the sign function.
For the ground-state in a harmonic trap, $\Psi_{F}$ is a Slater determinant  constructed with the single-particle eigenfunctions of a harmonic oscillator and the ground-state wavefunction $\Psi_0$ of the Tonks-Girardeau gas  takes the well-known form \cite{Girardeau01}
\beqa
\Psi_0(x_1,\dots,x_\N)=c_0\exp\left(-\frac{m\om_0}{2\hbar}\sum_{i=1}^\N x_i^2\right)\prod_{i<j}|x_{i}-x_{j}|, 
\label{GSTG}
\eeqa
where we note that $\psi_\nu=\prod_{i<j}|x_{i}-x_{j}|$ is a homogeneous function of degree $\nu=\N(\N-1)/2$, which yields $\sigma^2=\N^2/2$. Local correlation functions are identical in states  of  spin-polarized fermions $\Psi_{F}$ and the Tonks-Girardeau gas $\Psi_{TG}$ that are related by the Bose-Fermi duality (\ref{BFM}). As a result,   the value  $\sigma^2=\N^2/2$ applies to both systems.

{\it Calogero-Sutherland model.---}
The rational Calogero-Sutherland gas is described by the Hamiltonian
\begin{eqnarray}
 H=\sum_{i=1}^{\N}\left[\frac{p_i^2}{2m}+\frac{1}{2}m\omega_0^2 x_i^{2}\right]+\sum_{i<j}\frac{\lambda(\lambda-1)}{|x_{i}-x_{j}|^2} .
 \label{HCSM}
\end{eqnarray}
The ground state many-body wavefunction of the rational Calogero-Sutherland model takes the Bijl-Jastrow form \cite{Girardeau01}
\beqa
\Psi_0(x_1,\dots,x_\N)=c_0\exp\left(-\frac{m\om_0}{2\hbar}\sum_{i=1}^\N x_i^2\right)\prod_{i<j}|x_{i}-x_{j}|^\lambda,
\eeqa
which generalizes (\ref{GSTG}) for arbitrary $\lambda$. For $\lambda=0$, one recovers the one-dimensional ideal Bose gas, while for $\lambda=1$ the  Calogero-Sutherland model describes a Tonks-Girardeau gas.
We note that the function $\psi_\nu=\prod_{i<j}|x_{i}-x_{j}|^\lambda$  is a zero-energy eigenstate of the Hamiltonian in free space, Eq. (\ref{HCSM}) with $\om_0=0$,
\begin{eqnarray}
 H_{\rm free}=\sum_{i=1}^{\N}\frac{p_i^2}{2m}+\sum_{i<j}\frac{\lambda(\lambda-1)}{|x_{i}-x_{j}|^2}.
\end{eqnarray}
Indeed, 
\beqa
 H_{\rm free}\prod_{i<j}|x_{i}-x_{j}|^\lambda=0.
\eeqa
Further, $\psi_\nu$ is a homogeneous function of degree $\nu=\lambda\N(\N-1)/2$ and thus $\sigma^{2}=\nu+\N/2=\N[1+\lambda(\N-1)]/2$. This is consistent with the identity 
$\sigma^2=E(0)/(\hbar\om_0)$ as the ground-state energy of the trapped $\Psi_0$ is precisely $E(0)=\hbar\om_0\N[1+\lambda(\N-1)]/2$.

{\it Three-dimensional Unitary Fermi Gas.---} In a spin $1/2$ Fermi gas, the unitary regime can be reached via a Feshbach resonance tuning the contact interactions between spin-up and spin-down fermions making the interaction strength effectively divergent \cite{WernerCastin06}. In this regime, under harmonic confinement, the dynamics is scale invariant \cite{Castin04}.
Introducing hyperspherical coordinates $\vec{X}=(\vec{r}_1,\dots,\vec{r}_\N)$ with norm $X=\sqrt{\sum_j r_j^2}$ and  unit vector $\hat{n}=\vec{X}/X$, a low-energy state of a unitary Fermi gas in a harmonic trap has the structure \cite{WernerCastin06}
 \beqa
\Psi_0(\vec{X})=c_0e^{-\frac{m\om_0X^2}{2\hbar}}X^{\frac{E}{\hbar\om}-\frac{3\N}{2}}f(\hat{n}),
\eeqa
where $c_0$ is a normalization constant. Therefore, $\nu=\frac{E}{\hbar\om}-\frac{3\N}{2}$ and $\sigma^2=\frac{E}{\hbar\om}$.

\section{Experimental parameters}

In  what follows, we discuss the probe of QSL with ultracold gases in two different regimes, specifying a possible choice of the experimental parameters. We first consider hard-core bosons in the Tonks-Girardeau regime. To access it, experiments often  use  an optical lattice in a cross-dipole configuration which leads to an array of quasi-1D cigar-shaped clouds or tubes  \cite{Paredes04,Kinoshita04,Palzer09}. The typical number of particles per tube is $N=50$ and the values of the axial width  reported are in the range $10-50$ $\mu$m, with the dynamics being probed in $100-2000$ $\mu$s. 

Bose-Fermi duality guarantees that the QSL for a quantum state in the Tonks-Girardeau regime is identical to that of the corresponding state in the dual system, the polarized Fermi gas in one spatial dimension. As a result, an alternative platform to study QSL in ultracold gases is that presented in \cite{Zurn12,Wenz13}.
In this case, typical experimental conditions involve cigar-shaped traps  with a longitudinal frequency $\om_0/2\pi=1$ kHz and a transverse frequency an order of magnitude higher, $\om_{\perp}/2\pi=10$ kHz. 
The characteristic time scale for the cloud size to grow is largely set by the inverse of $\om_0$. For instance, in a time of flight experiment, $b(t)=\sqrt{1+\om_0^2t^2}\approx \om_0t$, for $t\gg\om_0^{-1}$.
The role of the number of atoms can be studied at the few-body level by varying $N$. The range $N=1,\dots, 5$ was reported in \cite{Wenz13}.

As an additional platform, we consider the 3D unitary Fermi gas. This is often realized in an anisotropic trap in which isotropic expansions can be engineered \cite{Deng16,Deng18}. For instance, Deng et al.  \cite{Deng18sci} reported a modulation of the trap frequency from the initial value $\om_x(0)/2\pi=1200$ Hz  
 to the final value $\om_x(t_F)/2\pi=300$ Hz, keeping  constant the other frequencies $\om_y(0)/2\pi=\om_z(0)/2\pi=300$ Hz, in the span of $800$ $\mu$s. In this process the expansion factor varies form $b(0)=1$ to about $b(t_F)=3$. 
Actually, the experiment  \cite{Deng18sci}  determined the nonadiabatic scaling factor $Q^*$ from time-of-flight imaging, which is needed to reconstruct the integrated energy fluctuations $\gamma(\tau)$ in this proposal.  A similar experimental setting with an isotropic trap  would  constitute a natural platform to explore QSL following this proposal.

\section{Relevant Hilbert Spaces}

The main body of the manuscript provides a  protocol to the determine  distances between quantum states and the speed of evolution (energy fluctuations) by well-established measurements of the density profile, routinely used in ultracold atom laboratories. Ultracold gases are highly dimensional continuous-variable systems of many particles on which standard approaches to quantum state tomography and fidelity estimation are not feasible in the laboratory. 
Indeed, a typical setting for state tomography involves states of $\N$ qubits in the Hilbert space  $\mathcal{H}=\mathbb{C}^{2\N}$,  with rather limited values of $\N$. 

Our method is scalable to an arbitrary number of particles. Consider an atom-number state in spatial dimension $D$. Being the system of continuous variables, the Hilbert space of a single-particle is infinitely dimensional and spanned by the set of square-integrable functions 
\begin{equation}
\mathcal{L}^2(\mathbb{R}^D,d\vec{r})=\left\{|\Psi\rangle:\int_{\mathbb{R}^D}|\langle \vec{r}|\Psi\rangle|^2d\vec{r}<\infty\right\}.
\end{equation}

Mathematically, our proposal provides an experimental (and theoretical) approach to determine fidelities, Bures distances, speed of evolution (energy fluctuations) and speed limits in a Hilbert space
\begin{equation}
\mathcal{H}=\bigotimes_{n=1}^\N\mathcal{L}^2(\mathbb{R}^D,d\vec{r}_i)\cong\mathcal{L}^2(\mathbb{R}^{\N D}),
\end{equation}
this is, the space of square-integrable wavefunction in spatial dimension $\N D$.
Bose and (spin-polarized) Fermi systems live in the corresponding symmetric and antisymmetric subspaces
\begin{eqnarray}
\mathcal{H}_{\rm bosons}&=&\mathcal{S}_+\mathcal{H}={\rm Sym} (\mathcal{H})\\
\mathcal{H}_{\rm fermions}&=&\mathcal{S}_-\mathcal{H}=\bigwedge_{n=1}^\N\mathcal{L}^2(\mathbb{R}^D),
\end{eqnarray}
defined by the action  of the symmetrizing and anti-symmetrizing operators  $\mathcal{S}_\pm$ to ensure Bose and Fermi statistics. 
We note that these Hilbert spaces  are  infinite-dimensional. 

\section{Dilatation Operator and Moments of the Squeezing Operator}

The many-particle squeezing operator is defined as
\beqa
C=\frac{1}{2}\sum_{i=1}^\N (\vec{r}_i\cdot\vec{p}_i+\vec{p}_i\cdot\vec{r}_i)=-i\hbar\frac{\N D}{2}-i\hbar \sum_{i=1}^\N\vec{r}_i\cdot\nabla_i.
\eeqa
It acts as the generator of dilatations described by the unitary
\beqa
T_{\rm dil}&=&\exp\left[-i\frac{\log b}{\hbar}C\right].
\eeqa
In the coordinate representation,
\beqa
T_{\rm dil}\Psi\left(\vec{r}_1,\dots,\vec{r}_\N\right)=b^{-\frac{\N D}{2}}\Psi\left(\frac{\vec{r}}{b},\dots,\frac{\vec{r}_\N}{b}\right).
\eeqa
Consider a quantum state of the form
\beqa
\label{Psi0}
\Psi_0(\vec{r}_1,\dots,\vec{r}_\N)=c_0e^{-\frac{m\om_0}{2\hbar}\sum_{i=1}^\N \vec{r}_i\,^2}\psi_\nu(\vec{r}_1,\dots,\vec{r}_\N),
\eeqa
where $c_0$ is a normalization constant
and $\psi_\nu(\vec{r}_1,\dots,\vec{r}_\N)$ satisfies
\beqa
\left(\sum_{i=1}^\N \vec{r}_i\cdot\nabla_i\right) \psi_\nu(\vec{r}_1,\dots,\vec{r}_\N)=\nu\psi_\nu(\vec{r}_1,\dots,\vec{r}_\N).
\eeqa
One then finds
\beqa
T_{\rm dil}\Psi_0(\vec{r}_1,\dots,\vec{r}_\N)=b^{-\frac{\N D}{2}-\nu}c_0e^{-\frac{m\om_0}{2\hbar b^2}\sum_{i=1}^\N \vec{r}_i\,^2}\psi_\nu(\vec{r}_1,\dots,\vec{r}_\N).
\label{TdilPsi0}
\eeqa
To determine the expectation value of the $k$-th moment of $C$, let us introduce the generating function \cite{Beau20}
\beqa
A_C(b)=\la \Psi_0|T_{\rm dil}\Psi_0\ra,
\eeqa
in terms of which 
\beqa
\la \Psi_0|C^k|\Psi_0\ra&=&\left(-i\hbar b\frac{d}{db}\right)^kA_C(b)\big|_{b=1}. 
\eeqa
Explicit evaluation of $A_C(b)$ for a state of the form (\ref{Psi0}) is possible using  (\ref{TdilPsi0})  to rewrite the multidimensional integral in terms of the normalization constant $c_0$. Without requiring explicit knowledge of $\psi_\nu(\vec{r}_1,\dots,\vec{r}_\N)$, one  finds
\beqa
A_C(b)=\left[\frac{b}{2}\left(1+\frac{1}{b^2}\right)\right]^{-\sigma^2},
\eeqa
whence it follows that
\beqa
\la \Psi_0|\hat{C}^2|\Psi_0\ra&=&\hbar^2\sigma^2,
\eeqa
with $\sigma^2=\nu+\frac{D\N}{2}$.

\section{QSL in the Adiabatic Limit}
The adiabatic limit of QSL requires some care as the instantaneous energy dispersion, being of order $\mathcal{O}(1/\tau)$, is suppressed as $\tau\rightarrow\infty$.
However, the integrated energy fluctuations do not vanish. 
To analyze the adiabatic limit of 
\beqa
\gamma(\tau)=\sigma\int_0^\tau dt \sqrt{\om(t)^2 \left[\left(Q(t)^*\right)^2-1\right]},
\eeqa
we first substitute the adiabatic solution of the Ermakov equation
\beqa
b(t)=\sqrt{\frac{\om_0}{\om(t)}}
\eeqa
in the expression of $Q^*$ and find
\beqa
Q^{\ast}(t)&=&\frac{\om_0}{\om(t)} \left[\frac{1}{2b(t)^{2}}+\frac{\om(t)^2 b(t)^2}{2\om_0^2}+\frac{\dot{b}(t)^2}{2\om_0^2}\right]\nonumber\\
&\approx&1+\frac{\dot{b}(t)^2}{2\om_0\om(t)}\nonumber\\
&=&1+\frac{\dot{\om}(t)^2}{8\om(t)^4}.
\eeqa
This adiabatic value of $Q^{\ast}(t)$ agrees with that under transitionless quantum driving \cite{Beau16,AbahLutz18,Beau20}.
Noting that
\beqa
\left(Q(t)^*\right)^2-1=\frac{\dot{\om}(t)^2}{4\om(t)^4}+\frac{\dot{\om}(t)^4}{64\om(t)^8},
\eeqa
the length of the path travelled under slow driving reads 
\beqa
\gamma(\tau)\approx \sigma\int_0^\tau dt \sqrt{\frac{\dot{\om}(t)^2}{4\om(t)^2}+\frac{\dot{\om}(t)^4}{64\om(t)^6}}.
\eeqa
To leading order, assuming $\om(s)$ monotonic, one finds
\beqa
\gamma(\tau)= \sigma\alpha\log\left(\frac{\om(\tau)}{\om_0}\right)^{\frac{1}{2}} +\mathcal{O}(1/\tau),
\eeqa
where $\alpha={\rm sgn}(\dot{b})$. This agrees with the result under transitionless quantum driving, Eq. (23) in the main text.
The geodesic $\mathcal{L}(\tau)$ depends only on the initial and final states and makes no reference to the actual dynamics. 
As a result,  the behavior of $\delta{\cal L}(\tau)$ under transitionless quantum driving is equal to that in the adiabatic limit.

\section{Time-dependent Frequencies by Reverse Engineering and Transitionless Quantum Driving}

The engineering of controlled expansions and compression of ultracold gases in the laboratory has been discussed in a number of works. The reverse engineering of the self-similar scaling dynamics \cite{Chen10,delcampo11} is summarized here to make the manuscript self-contained.

Traditionally, the scaling law Eq. (8) in the main body of the manuscript 
\beqa
\label{psitSM}
\Psi\left(t\right)&=&
\frac{1}{b^{\frac{D\N}{2}}}\exp\left[i\frac{m\dot{b}}{2\hbar b}\sum_{i=1}^\N \vec{r}_i\,^2-i\int_{0}^t\frac{E(0)}{\hbar b(t')^2}dt'\right]\Psi\left(\frac{\vec{r}_1}{b},\dots,\frac{\vec{r}_\N}{b},t=0\!
\right)\,,
\eeqa
describes the evolution of a quantum state, following a modulation of the trapping frequency $\omega(t)$. To that end, one solves the Ermakov equation $\ddot{b}+\omega(t)^2b=\omega_0^2/b^3$ as an initial value problem subject to the initial conditions $b(0)=1$ and $\dot{b}=0$ to determine the scaling factor $b(t)$ as a function of time, which is required to specify the solution (\ref{psitSM}) of the time-dependent Schr\"odinger equation.

\begin{figure}[t]
\includegraphics[width=1\linewidth]{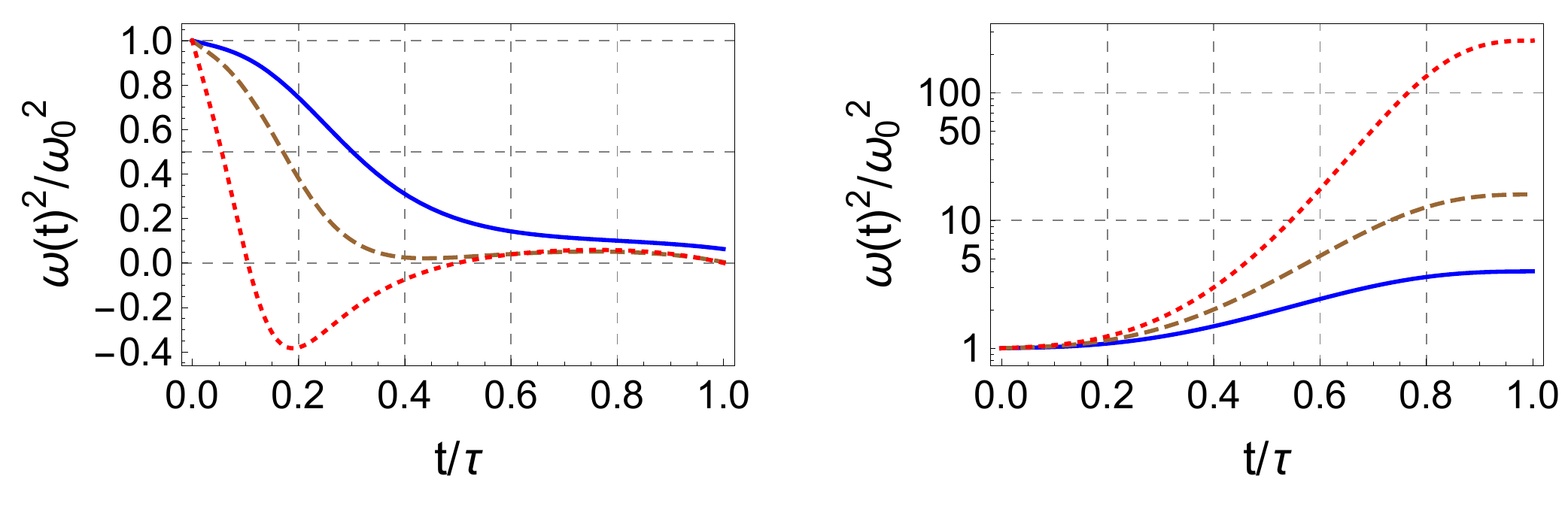}
\vspace{-0.5cm}
\caption{\label{FigSM1} {\bf  Frequency driving by reverse engineering the scale-invariant dynamics}. In an expansion (left), for a fixed duration of the protocol $\tau=10\om_0^{-1}$, the square of the driving frequency is shown as  a function of time for different values of the final frequency $\om_F/\om_0=1/2,1/4,1/16$ from top to bottom. Smaller values of the final frequencies imply a higher expansion factor $b(\tau)=\sqrt{\om_0/\om_F}$ and  may require the transient inversion of the trap associated with values $\om^2(t)<0$. However,  fast expansion protocols violating the adiabaticity condition $\dot{\om}/\om^2\ll 1$ can be engineered with  $\om^2(t)>0$  at all times, without the need for inverting the trap. In a compression of the trap (right), it is found that $\om^2(t)>0$  even for very fast protocols, as shown for $\om_F/\om_0=2,4,16$ from bottom to top, in a logarithmic scale, for clarity.  
}
\end{figure}

Reverse engineering proceeds in the opposite way. It first identifies a time dependence of interest for the scaling factor $b(t)$.
Such trajectory must satisfy some boundary conditions for the time dependent state $\Psi(t)$ to reduce to a stationary state at time $t=0$. In particular,
\beqa
b(0)=1,\quad \dot{b}(0)=0,
\label{BC1}
\eeqa
where the latter condition guarantees the vanishing of the oscillatory phase in Eq. (\ref{psitSM}).
Similarly, for the state $\Psi(t)$ to become an energy eigenstate of the final Hamiltonian $H(\tau)$, it must be the case that
\beqa
b(\tau)=\sqrt{\frac{\om_0}{\om(\tau)}},\quad \dot{b}(\tau)=0.
\eeqa
These boundary conditions can be supplemented by additional ones of the form 
\beqa
\ddot{b}(0)=\ddot{b}(\tau)=0,
\label{BC3}
\eeqa
to guarantee the smoothness of an interpolating ansatz for $b(t)$.
Consider for instance the polynomial ansatz 
\beqa
b(t)=\sum_{n=0}^5c_n(t/\tau)^n.
\label{PAnsatz}
\eeqa
Imposing the boundary conditions (\ref{BC1})-(\ref{BC3}) in (\ref{PAnsatz}) suffices to fix the coefficients $c_n$, and readily yields  the result used in the text
\beqa
b(t)=1+10(t/\tau)^3(b(\tau)-1)-15(t/\tau)^4(b(\tau)-1)+6(t/\tau)^5(b(\tau)-1).
\label{Pbt}
\eeqa
This polynomial interpolation for the scaling factor is monotonic both for expansions and compressions.
Once the scaling factor is known, one proceeds to determine the frequency modulation $\om(t)$ that generates it by using the Ermakov equation as
\beqa
\om(t)^2=\om_0^2/b^{4}- \ddot{b}/b, 
\eeqa
plugging in the prescribed solution (\ref{Pbt}). It is in this sense that  the method reversely engineers the dynamics.  
The resulting expression for this polynomial ansatz reads
\beqa
\om(t)^2=\frac{\omega _0^4}{\left(\frac{(b(\tau) -1) t^3 \left(6 t^2+10 \tau ^2-15 \tau  t\right)}{\tau ^5}+1\right)^4}
-\frac{60 (b(\tau) -1) t \left(\frac{t}{\tau }-1\right) \left(\frac{2 t}{\tau }-1\right)}{\tau ^3\left(\frac{(b(\tau) -1) t^3 \left(6 t^2+10 \tau ^2-15 \tau  t\right)}{\tau ^5}+1\right)},
\eeqa
where $b(\tau)=\sqrt{\frac{\om_0}{\om(\tau)}}$ for short.
The explicit form of this frequency modulation has been discussed, for instance, in \cite{Chen10,delcampo11}. Figure \ref{FigSM1} shows the modulation as a function of time for several expansion and compression factors.
%

We next discuss the case of relevance to transitionless quantum driving. 
In the context of control theory and the engineering of STA, to specify the  modulation in time of the trap frequency one starts by identifying the desired initial and final trap-frequency values $\om_0$ and $\om_F$, as well as the duration of the process $\tau$.
An interpolating trajectory between the initial and final value can be chosen using an ansatz of the form
\beqa
\omega(t)=\omega(0)\sum_{n=0}^5c_n(t/\tau)^n.
\eeqa
The coefficients can be determined by imposing boundary conditions such as
\beqa
\omega(0)&=&\omega_0, \quad \omega(\tau)=\omega_F, \\
\dot{\omega}(0)&=&0, \quad \dot{\omega}(\tau)=0, \\
\ddot{\omega}(0)&=&0, \quad \ddot{\omega}(\tau)=0. 
\eeqa
If one wishes to choose a higher order polynomial, coefficients can be fixed by imposing the vanishing of higher-order derivatives.
The interpolating ansatz satisfying the above boundary conditions is 
\beqa
\omega(t)=\omega_0+(\omega_F-\omega_0)\left[10(t/\tau)^3-15(t/\tau)^4+6(t/\tau)^5\right].
\eeqa
This $\omega(t)$ can be chosen as a reference modulation of the trap frequency. Only for slow driving, whenever the adiabaticity  condition $\dot{\om}/\om^2\ll1$ satisfied, the dynamics is adiabatic.
In this case, the evolution of an initial state $\Psi(0)=\Psi_0$ follows the instantaneous ground-state at all times.
Transitionless quantum driving guides the dynamics through the same reference trajectory in processes of duration $\tau$ short enough as to break the adiabaticity condition $\dot{\om}/\om^2\ll1$. 
The scaling factor is still guaranteed to be
\beqa
b(t)=\sqrt{\frac{\om_0}{\om(t)}},
\eeqa
which would correspond to the adiabatic solution of the Ermakov equation with $\om(t)$.
To do so, without the requirement for slow driving, an auxiliary counterdiabatic control term is included in the Hamiltonian. For expansions and compressions in scale invariant dynamics this term takes the form of the squeezing operator, given by Eq. (21) in the main body of the paper. Alternatively, in a moving frame associated with a canonical transformation, this additional term is removed and it suffices to modify the driving frequency  \cite{delcampo13cd,Deng18sci}.

\section{Error Propagation in the Analysis of Measurement Data}

Any analysis of TOF data requires consideration of the  uncertainty in the data. There are three quantities of interest in the study of   QSL:
i)	Distance between quantum states (Bures distance), 
ii)	Speed of evolution (energy fluctuations), 
iii)	Minimum time given by the ratio of the distance over the velocity.
The complexity of determining the quantities i-iii) is essentially the same than that of measuring the nonadiabatic factor $Q^*$ which has already been done in the laboratory \cite{Deng18sci}.

Let us denote any of the quantities of interest i)-iii)   by $X(t)=X[b(t)]$. Note that their definition involves the function $b(t)$ (and its derivative), as measured  in the laboratory. Specifically, let us assume that experimentally, measurements of $b(t)$ are collected at discrete times of evolution $t_m=m\tau/M$  varying from $t_0=0$ to $t_M=\tau$ as $m=0,\dots M$. 
We assume that the experimental data for $b_m=b(t_m)$ has standard deviation $s_{b_m}$. Let us use a discretization of the derivative, e.g., $\dot{b}(t_m)=(b_{m+1}-b_m)M/\tau$ (or any other favorite discretization including $\{b_m,b_{m\pm 1}\}$). 
One is  then  left with the problem of estimating the uncertainty of the quantity $X=X(\{b_m\})$.
Assuming that uncertainties at different times $t_m$ are uncorrelated (e.g. if resulting from pixelation),
one can  use the error propagation formula to determine the uncertainty in the  quantity $X$
\begin{eqnarray}
s_X=\sqrt{\sum_{m=0}^M\left(\frac{dX}{db_m}\right)^2s_{b_m}^2}.
\end{eqnarray}

One may  wonder whether the propagation of errors in the estimated quantities blows up with the time of evolution, restricting the proposal to short times of evolution.
Clearly this is not the case. Indeed, note that $\mathcal{L}(t_m)$ is local in time in the sense that it depends only on $\{b_m,b_{m\pm 1}\}$ through the discretized form of the time derivative $\dot{b}$ . Similarly, the speed of evolution is the time-average energy fluctuation and even if it includes  all uncertainties $s_{b_m} (m=0,\dots,M)$ it is normalized by the total time $\tau$.

\section{Time-of-flight Imaging}

Let us consider that  the process of interest occurs in the time interval $[0,\tau]$ and is associated with the  nonadiabatic dynamics resulting from an arbitrary modulation of $\omega(t)$.
At any intermediate time $0\leq t_m\leq\tau$, TOF imaging is used to determined $b_m=b(t_m)$ at $t_m$.
To describe the complete evolution including the interval $[0,t_m]$ and the subsequent TOF  for $t>t_m$, we exploit the unitarity of the dynamics. The time-evolution operator is given by the group property
$U(t,0)=U_{\rm TOF}(t,t_m)U(t_m,0)$.
At any time $t_m$, $\Psi(t_m)=U(t_m,0)\Psi(0)$ is given by equation (8) in the manuscript. 
Such state is generally a nonequilibrium state that can be expressed as a coherent quantum superposition of the energy eigenstates $\Phi_n(t_m)$ of the Hamiltonian $H(t_m)$ at $t_m$:
\beqa
\Psi(t_m)=\sum_n c_n(t_m)\Phi_n(t_m),\quad  c_n(t_m)=\la \Phi_n(t_m)|U(t_m,0)\Psi(0)\ra.
\eeqa
As the dynamics is scale invariant, the subsequent TOF evolution until time $t=t_m+t_{\rm{TOF}}$ of each energy eigenstate is given, in analogy  with equation (8), by
\beqa
\Phi_n(t)=
\frac{1}{b_{\rm{TOF}}^{\frac{D\N}{2}}}\exp\left[i\frac{m\dot{b}_{\rm{TOF}}}{2\hbar b_{\rm{TOF}}}\sum_{i=1}^\N \vec{r}_i\,^2-i\int_{0}^{t_{\rm{TOF}}}\frac{E_n(t_m)}{\hbar b_{\rm{TOF}}(t')^2}dt'\right]
 \Phi_n\left(\frac{\vec{r}_1}{b_{\rm{TOF}}},\dots,\frac{\vec{r}_\N}{b_{\rm{TOF}}},t_m\!\right)\,.
\eeqa
As TOF imaging is associated with the sudden released of the trapped system, by setting $\om(t)=0$ for all $t\geq t_m$, the scaling factor associated with the TOF expansion is 
\beqa
b_{\rm{TOF}}=\sqrt{1+\omega(t_m)^2t_{\rm{TOF}}^2}.
\eeqa
Taking $\Psi(t_m)$ as the initial state for the TOF expansion,
\beqa
\Psi(t)=\sum_n c_n(t_m)\Phi_n(t).
\eeqa
Let us focus on the  particle dispersion in the position representation that governs the average square radius of the atomic cloud
\beqa
R^2(t)&=&\la\Phi_n(t)| \sum_{i=1}^\N \vec{r}_i\,^2|\Phi_n(t)\ra\\
&=&b_{\rm{TOF}}^2\la\Phi_n(t_m)| \sum_{i=1}^\N \vec{r}_i\,^2|\Phi_n(t_m)\ra\\
&=&b_{\rm{TOF}}^2R^2(t_m),
\eeqa
where we have exploited scale invariance [In addition, it further holds that $R^2(t_m)=b(t_m)^2R^2(0)$].

Above, we have implicitly assumed that TOF expansion occurs in the same spatial dimension $D$ as the dynamics in $[0,t_m]$.
When the dynamics under study in the interval $[0,t_m]$ is strongly confined along one or two axes with frequency $\om_{\perp}\gg\om(t_m)$ (for any $t_m\in[0,\tau]$),  a high potential energy is stored in the confined degrees of freedom. When suddenly switching off the whole confining potential for TOF imaging, the short-time dynamics is governed by the expansion along these degrees of freedom and can be effectively negligible along the remaining ones. As a result, after the time scale $\om_{\perp}^{-1}$ interparticle interactions become negligible and the time evolution is already generated by a purely kinetic Hamiltonian, which ensures that the TOF expansion remains scale invariant, i.e., self-similar.


\end{document}